\documentclass[times,authoryear]{elsarticle}

\usepackage{jasr}
\usepackage{framed,multirow}

\usepackage{graphicx}
\usepackage{xcolor} 
\usepackage{dcolumn}
\usepackage{bm}
\usepackage{ulem} 
\usepackage{graphics}
\usepackage{amssymb,amsmath}

\usepackage[switch]{lineno}

\definecolor{myblue}{rgb}{0.05,0.1,0.5}

\definecolor{myred}{rgb}{0.5,0.05,0.1}

\usepackage[citebordercolor=white]{hyperref}

\journal{Advances in Space Research}

\begin{document}
\verso{K.~Dolgikh \textit{etal}}
\begin{frontmatter}

\title{Images of the Ultra-High Energy Cosmic Rays from Point Sources}


\author[1,4]{Konstantin \snm{Dolgikh}\corref{cor1}}
\cortext[cor1]{Corresponding author: 
  e-mail: dolgikh.ka15@physics.msu.ru;
}
\ead{dolgikh.ka15@physics.msu.ru}
\author[2]{Alexander \snm{Korochkin}}
\ead{alexander.korochkin@ulb.be}
\author[1]{Grigory \snm{Rubtsov}}
\ead{rgbeast@yandex.ru}
\author[3]{Dmitry \snm{Semikoz}}
\ead{dmitri.semikoz@apc.univ-paris7.fr}
\author[1]{Igor \snm{Tkachev}}
\ead{tkachev@ms2.inr.ac.ru}

\affiliation[1]{organization={Institute for Nuclear Research of the Russian Academy of Sciences},
                addressline={avenue of the 60th anniversary of October, 7A},
                city={Moscow},
                postcode={117312},
                country={Russia}}

\affiliation[2]{organization={Université Libre de Bruxelles},
                addressline={CP225 Boulevard du Triomphe},
                city={Brussels},
                postcode={1050},
                country={Belgium}}

\affiliation[3]{organization={APC, Universit\'e Paris Cit\'e, CNRS/IN2P3, CEA/IRFU},
                addressline={Observatoire de Paris},
                city={Paris},
                postcode={119 75205},
                country={France}}

\affiliation[4]{organization={Lomonosov Moscow State University, Faculty of Physics},
                addressline={Leninskie gory, 1},
                city={Moscow},
                postcode={119991},
                country={Russia}}

\received{7 December 2023}
\finalform{25 June 2024}
\accepted{20 June 2024}
\availableonline{ June 2024}
\communicated{K. Dolgikh}

\begin{abstract}
Our latest paper \citep{dolgikh2022causticlike} investigates the effects of UHECR propagation in a turbulent intergalactic magnetic field in the small-angle scattering regime, specifically focusing on the non-trivial caustic-like pattern that arises with strong deviation from isotropy. In this paper, we explore the effect of the observer's position on the measurement of source flux  at a given distance. We examine three types of source locations, characterized by the density of cosmic rays from a given source at the observation point, which we call {magnetic} knots, {magnetic} filaments and {magnetic} voids. We also investigate the energy spectrum in these different cases and present simulated images of the source as it appears on the observer's telescope after propagation in the combination of intergalactic and {Galactic} magnetic fields. We show that hot spots in the UHECR data can arrive due to combined distortions of {the} source images on the intergalactic and {Galactic} magnetic fields. Also the fact that flux of most nearby sources is diluted in the {magnetic} voids affects source population studies.
\end{abstract}


\end{frontmatter}


\section{Introduction}

Contrary to photons and neutrinos, Ultra-High Energy Cosmic Rays (UHECR) are charged particles and they are unavoidably deflected in the Galactic and intergalactic magnetic fields {(IGMF)}. Expected large number of UHECR sources in the Universe together with large deflections of UHECR from source positions prevents {us from doing} astronomy except for the highest energies $E>6\cdot 10^{19}$~eV, where the spectrum has {a} cutoff due to interactions with {the} Cosmic Microwave Background (CMB). This {so-called} GZK cutoff was predicted by Greizen, Zatsepin and Kuzmin in 1966 \citep{Greisen:1966jv,Zatsepin:1966jv}. {Both original papers \citep{Greisen:1966jv} and \citep{Zatsepin:1966jv} provide calculations of pion photoproduction by the proton primaries while at the same time point to the effect of photodisintegration which takes place in {the} case of primary nuclei at nearly the same energy, see~\citet{Kampert:2012vi} for historical review.}

{The cutoff in the cosmic ray spectrum was observed 40 years later by {the High Resolution Fly's Eye (HiRes)} experiment in 2007 \citep{HiRes:2007lra} and was confirmed shortly thereafter with higher exposure by the Pierre Auger Observatory (Auger) in 2008~\citep{PierreAuger:2008rol}. Later in 2013 the cutoff was supported by the Telescope Array (TA) results~\citep{TelescopeArray:2012qqu}.}  {Recent Auger {measurements} show that the observed spectrum suppression may be {the} result {of} the combination of the impact of the maximum possible acceleration energy of heavy nuclei in the sources and the GZK effect~\citep{PierreAuger:2020kuy}.} 
{In certain models, see e.g.~\citet{Eichmann:2022ias}, the ultra-high energy part of the cosmic ray spectrum is dominated with the local sources. In the latter case the maximum energy of acceleration determines the maximum energy of the cosmic rays.}
{The} existence of {the} GZK {effect} guarantees that UHECRs with energies $E>6\cdot 10^{19}$~eV come from {the} local Universe with significant contribution of sources located in nearby Large Scale Structures (LSS) with {a} distance to our Milky Way galaxy $R<200$ Mpc. Since {the} LSS at those distances is significantly non-uniform, this gives a hope to see individual  UHECR sources. Astronomy with UHECR{s} at the highest energies $E>6\cdot 10^{19}$~eV was one of the main motivations for the construction of {the} Auger and TA experiments.

For cosmic ray protons deflection in the $B_\mathrm{Gal} \sim \mu$G {Galactic} magnetic field is only {a} few degrees for $E>6 \cdot 10^{19}$~eV  depending on {the} direction. {At the same time even the strongest possible IGMFs with the strength of the} order of $B_\mathrm{IGMF} \leq n$G are again not efficient to disturb {the} UHECR proton directions from point sources. Thus one expect{s} to see point or slightly extended sources of protons $E>6\cdot 10^{19}$~eV in UHECR data. However, even with {a} large exposure Auger and TA have not found any signature of point sources. This observation is consistent with {the} statement made by Auger that most of {the} UHECR{s} with $E>6\cdot 10^{19}$~eV are intermediate and heavy nuclei. In the case of heavy nuclei, the {Galactic} magnetic field disturbs and even washes out images of UHECR sources~\citep{Giacinti:2010dk,Giacinti:2011uj}.

After years of observation, Auger and TA see several anomalies in the sky that look like hot spots with a radius of 10-20 degrees. These hot spots can be images of nearby point sources, disturbed both in Galactic and extragalactic magnetic fields.

Previous studies have explored the effects of intergalactic magnetic fields on the trajectories of UHECRs. One such study, \citet{Y_ksel_2012}, investigated the Centaurus A UHECR excess and its potential connection to the local extragalactic magnetic field. This study shows that, assuming a proton composition, the strength of the extragalactic magnetic field should be at least 20 nG. {This value} value is in conflict with existing observational constraints \citep{Pshirkov:2015tua,Katz:2021iou,Jedamzik:2018itu}. In {the} present work we assume a high but realistic magnetic field strength.

{The} main goal of the present study is to show that the observed 10-20 degree radius hot spots in {the} UHECR data can be signals from nearby point sources, disturbed both in {the} Galactic and intergalactic magnetic field. In particular, we show examples of sources, which can explain {the} TA hot spot. We show that {a} caustic-like pattern {arising after propagation through the} intergalactic magnetic field amplif{ies} small number of sources in filaments and knots and at the same time reduce{s the} contribution of most of {the} nearby sources.  

\section{Simulation}
To obtain images and spectra of {the} UHECR sources, we rely on numerical simulations of {the} UHECR propagation. Our simulation procedure generally consists of two steps. In the first step, we propagate UHECRs in {the} turbulent intergalactic magnetic fields, tracing their paths until they reach the edge of our galaxy. In the second step, we consider the effect of the {Galactic} magnetic field and construct images at the Earth's position.

For intergalactic propagation we use 3D Monte Carlo code \texttt{CRbeam}\footnote{\url{https://github.com/okolo/mcray}}\citep{Kalashev_2023} which allows to take into account UHECR deflections in magnetic fields and interactions with {the} CMB and {the extragalactic background light (EBL)}. The general setup of {these} simulations is the same as in our previous paper \citet{dolgikh2022causticlike}. {Namely} {we consider the source of UHECRs surrounded with the homogeneous turbulent magnetic field with a fixed strength and correlation length. The main difference is that the source emits UHECRs not isotropically, but in the} form of the directional cone with a given opening angle and an axis direction (thus isotropic source is a special case with {the} opening angle equals $\pi$). In all simulations we keep flat profile of the cone. Another important modification is that we catch particles when they hit a small spherical observer which represents our galaxy. The radius of observer was chosen to be 100, 200 or 1000 kpc, the last one for {magnetic} void cases. Once the particle hits the observer, we record its energy, momentum, and position on the sphere. We subsequently use this information to reconstruct the source image. 

\texttt{CRbeam} represents turbulent magnetic field with Kolmogorov's spectrum as a sum of plain waves with random directions and phases, following the method of \citet{Giacalone_1999}. The number density of waves is uniform in logarithmic scale by wavenumber. We fix the number of waves to be 463 with the largest eddy being 100 larger than the smallest. A smaller number of waves may lead to an unwanted effect when a wave with the longest wavelength dominates UHECR deflection. On the {other} hand, bigger number of waves increase calculation time. In our previous study \citet{dolgikh2022causticlike}, we carefully verified that existence of the caustic-like patterns in UHECR spacial distribution does not depend on either the method of generating the turbulent magnetic field or its spectrum. Therefore, we believe that our result are stable with respect to the technical parameters of the magnetic field generators such as number of modes and minimum scale of the {turbulence}. We keep the same random seed for magnetic field generation in all simulations within the paper and use \texttt{OpenMP} for multiprocessing acceleration. 

The magnetic field strength was set to $B=1$~nG and {the} correlation length $\lambda_\mathrm{c}=1$~Mpc. {The magnetic field with such $\lambda_\mathrm{c}$ may be produced as a result of the evolution of primordial IGMF with an initially scale-invariant power spectrum, as was shown in \citet{Mtchedlidze:2021bfy}. In most simulations we propagate charged particles with a rigidity of $R = 10$~EV. Since all the sources we are considering are at the distances less than 100 Mpc, and in most cases at distances of the order of 20-30 Mpc, we neglect interactions of UHECRs with the background fields. } 

{While the distances are small enough to neglect interactions, they are large compared to the size of the observer. Given the fact that the observer's sphere has the radius of tens and hundreds kpc and charged particles have nonnegligible deflections in IGMF with $B\propto1$~nG the problem of aiming at the observer arises. In practice this means that most of the simulated particles will miss the observer  thus increasing the amount of computational time spent in vain.} This is due to both a decrease in the angular size of the observer with distance {as viewed from the source} and to the fact that statistically more often the observer{s} find {themselves} in the region{s} of space where the cosmic ray flux from {the} source is lower than expected from standard inverse-square law \citep{dolgikh2022causticlike}. This prompts the need of {a} targeting algorithm {which could reduce the number of misses.} 

{This problem was already studied earlier. For example in \citet{CRPropaLearns} to increase the probability of hitting the observer the initial directions of particle momentum are drawn from the von Mises–Fisher distribution with parameters adjusted as the number of detected particles increases. Here we develop a new targeting algorithm to optimize the performance.}

{Our algorithm is iterative, and at each subsequent step, it heavily relies on the results obtained in the previous one. The essence of the algorithm lies in the fact that if we know the initial directions that led to hits from a distance to the observer $D_\mathrm{i}$, then we assume that the successful initial direction for slightly larger distance $D_\mathrm{i+1}$, will be close to the old ones. }

{The step \texttt{i} of the algorithm works as follows. The distance to the observer is set to $D_\mathrm{i}$ and the particles are emitted in the cone with the opening angle $\alpha_\mathrm{i}$ and axis direction $\theta_\mathrm{i}$, $\phi_\mathrm{i}$. The particles are emitted uniformly within the cone and the simulation stops once the number of particles hitting the target reaches some predefined value which we set to $N_\mathrm{hit} \approx 100$. After that the parameters of the cone are recalculated to be used in the next iteration. The new direction of the cone axis corresponds to the average initial direction of the captured particles, while the new opening angle is set to three times the maximum angle between the initial direction of the captured particles and the new cone axis. The rescaling factor of the cone opening angle and the distance increase step size (which was set to a fraction of the Larmor radius of the particle) were determined empirically and do not have strict mathematical justification. Eventually, the algorithm terminates when $D_\mathrm{i}$ reaches the desired distance to the source.}

{All the results presented in the this work (see Section \ref{sec:results}) were obtained using this algorithm, which significantly reduced computation time. The acceleration factor compared to the isotropic source can be roughly estimated as the ratio of the total solid angle to the solid angle of the final cone $4\pi/2 \pi (1-\cos(\theta))$. For most of our simulations, this ratio is of the order of 100.}

Finally{,} we take into account deflections in {the Galactic} magnetic field and obtain images of the sources as seen by {an} observer at {the} Earth's position inside the Galaxy. We use galactic magnetic lenses provided by CRPropa3 \citep{AlvesBatista:2016vpy, AlvesBatista:2022vem}, which enables efficient utilization of backtracking simulations to consider galactic deflections in forward simulations of UHECRs. {In particular, we adopt the lens calculated for the JF12 model of the Galactic magnetic field ('JF12full') \citep{AlvesBatista:2016vpy} with coherent, striated and turbulent components turned on \citep{JF_GMF_1,JF_GMF_2}. The turbulent component was assumed to have Kolmogorov spectrum with the correlation length of 60~pc and the strength that follows the JF12 turbulent field model \citep{JF_GMF_2}.}

\section{Results}\label{sec:results}
\subsection{Observed fluxes}
\begin{figure}
    \includegraphics[width=\linewidth]{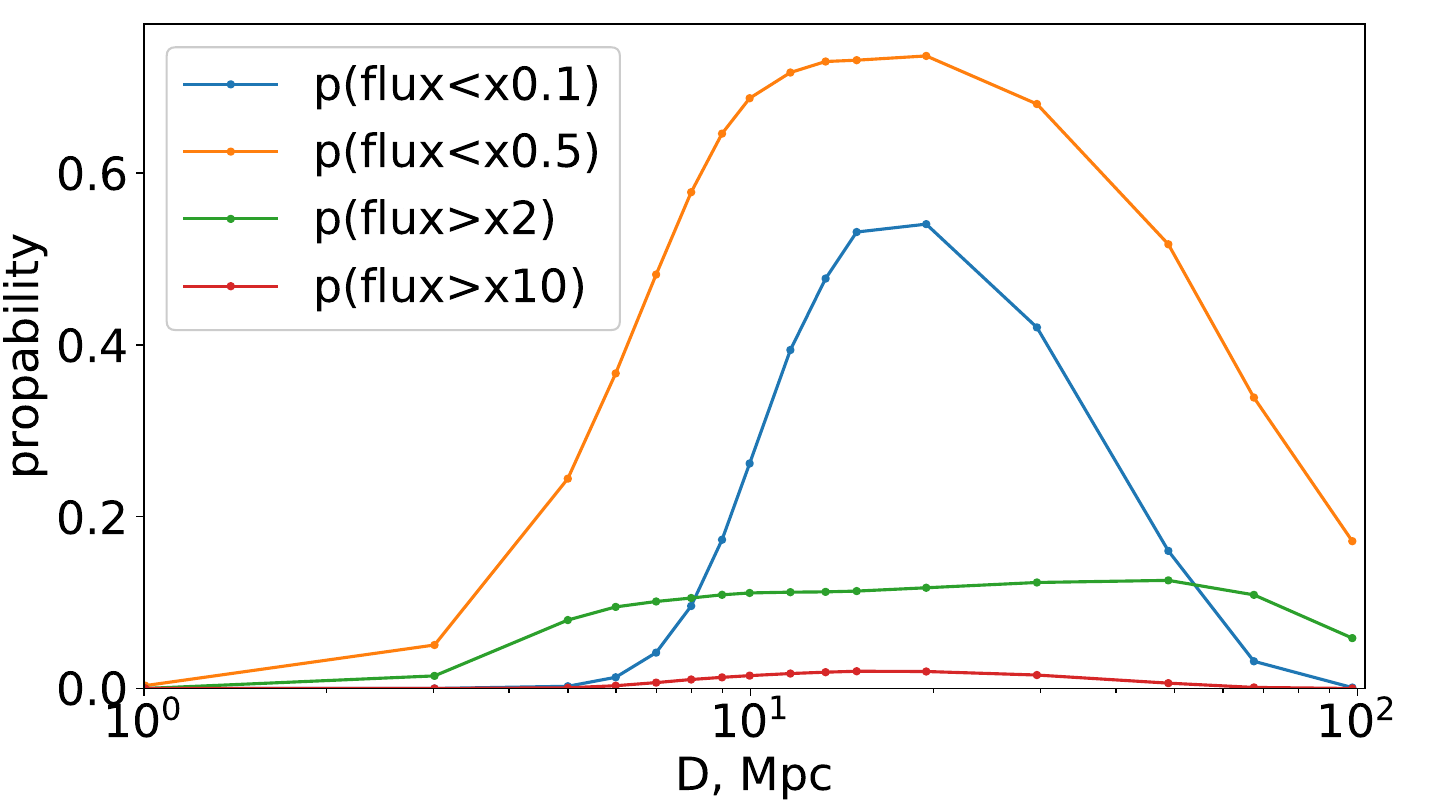}
    \caption{Probability of observer to be located in the region with flux higher or smaller compared to an average one as function of the distance to the source for {the particles} with {the rigidity of} $E=10$ EV.}
    \label{fig:propabilities}
\end{figure}

\begin{figure}
    \includegraphics[width=\linewidth]{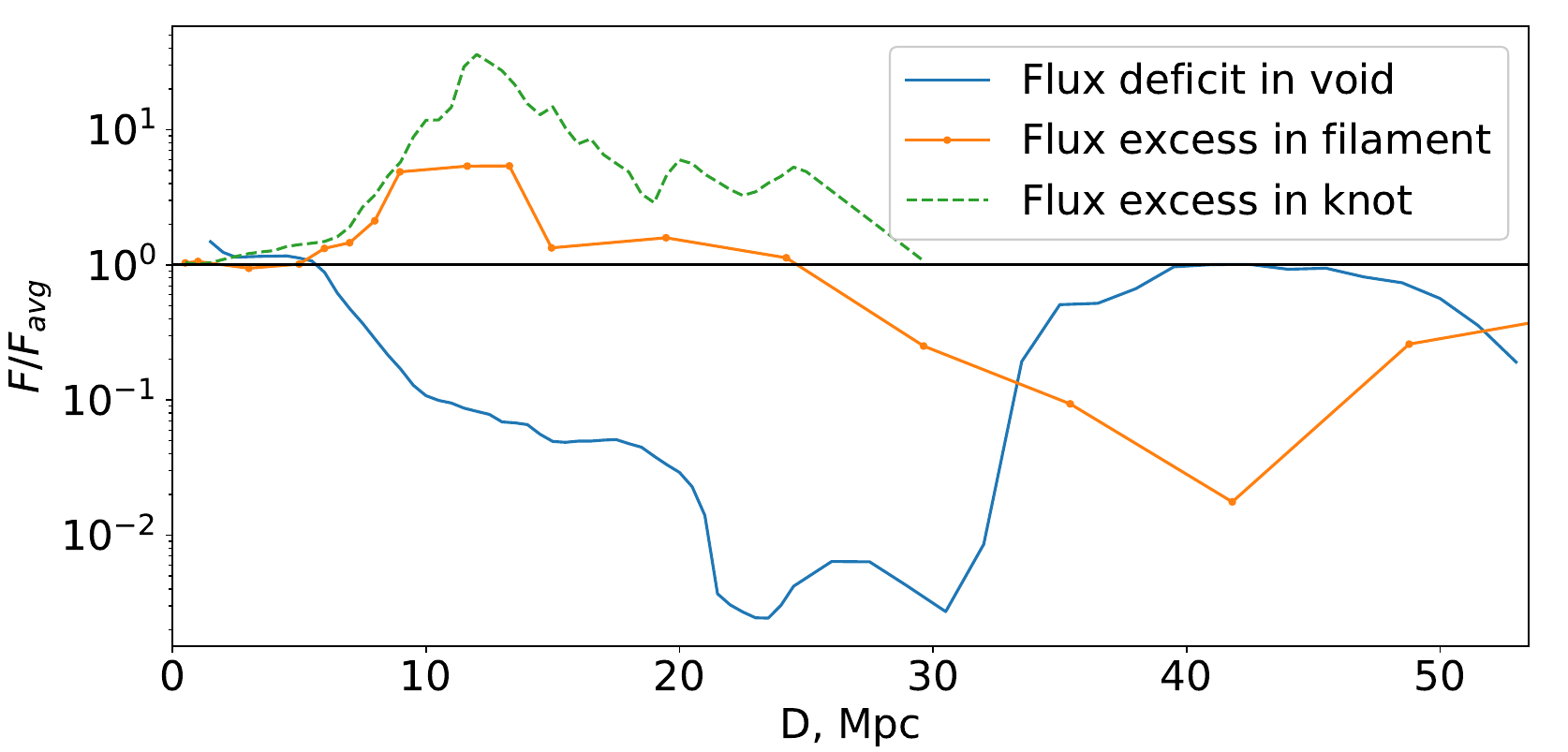}
    \caption{Example of the relative flux excess and deficit as function of the distance to the source for {the particles} with {the rigidity of} $E=10$ EV. {For this figure we define magnetic void, filament and knot at the distance of 10 Mpc.}}
    \label{fig:KnotDist}
\end{figure}

\begin{figure}
    \includegraphics[width=\linewidth]{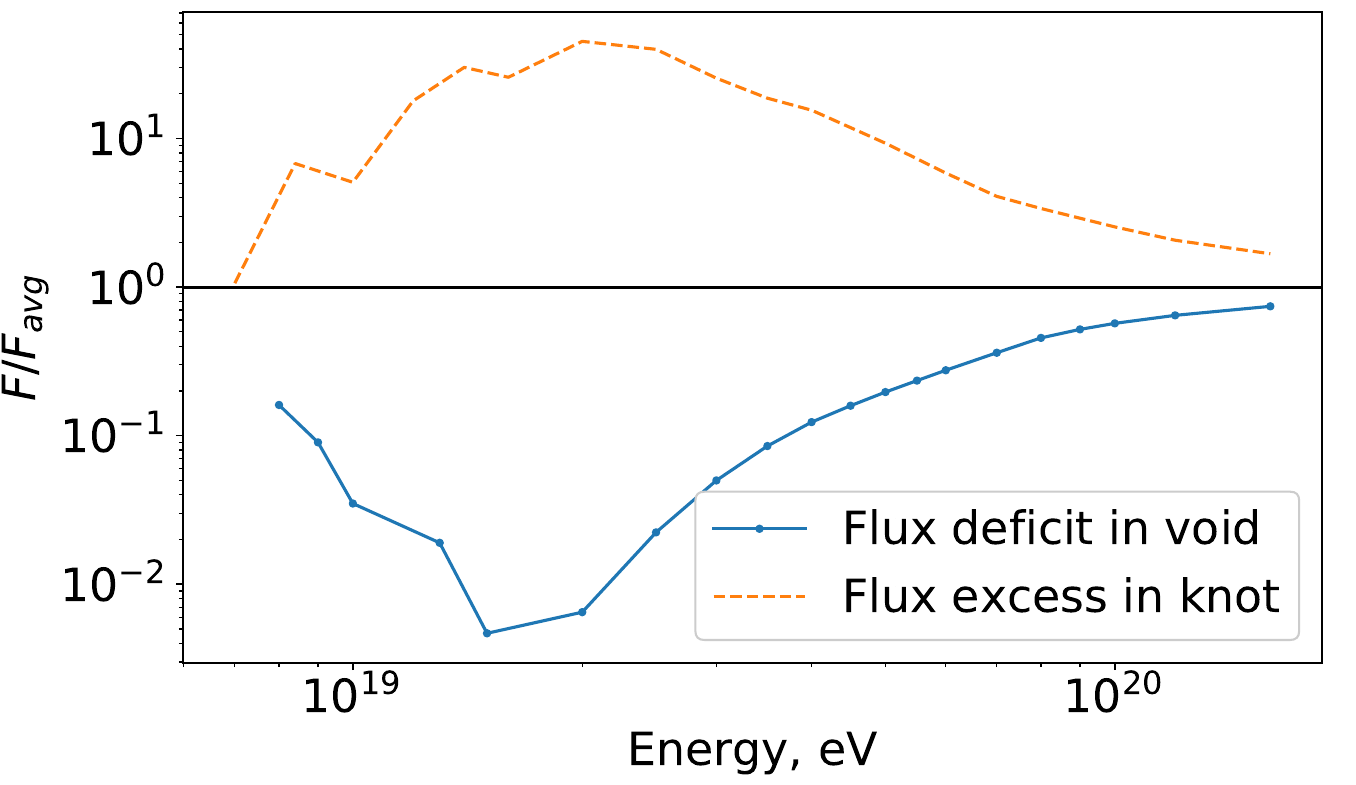}
    \caption{{Proton} energy spectrum of the source seen by an observer located in the knot at the distance $D=20$ Mpc and in the void at $D=30$ Mpc. Each spectral bin is normalized to the average flux at the same energy.}
    \label{fig:KnotSpectrum}
\end{figure}

In our previous paper \citet{dolgikh2022causticlike} we have shown that in the regime of small angle scattering UHECR from point source form a caustic pattern at distance equal to tens of coherent length from the source. This pattern consist of regions with density of UHECRs larger or smaller compared to isotropic averaged. {Due to the shape of these regions} on sphere around source we called them {magnetic} knots, filaments and voids. Typical flux in knots is 10 times larger compared to average, in filaments at least 2 times larger and in {magnetic} voids at least two times smaller. By average, we mean the expected flux from the source, assuming that it follows the inverse-square law.

We studied propagation of UHECR protons with energies $E>10^{19}$ eV in the turbulent intergalactic magnetic field with strength $B=1$~nG and coherent length $\lambda_\mathrm{c}=1$~Mpc. For such parameters of magnetic field non-trivial UHECR pattern exist at all distances in nearby LSS up to 100 Mpc, from which most of UHECR at highest energies arrive in the Earth detectors.

First, we examine the properties of the observed signal as a function of UHECR energy and a distance to the source. In Fig. \ref{fig:propabilities} we show probability that at a given distance from the source UHECR flux differs from an average one in 2 or 10 times. One can see, that most probably the flux from the given source will be lower compared to the average one. For 10 \% of sources it is at least two times higher while in up to 75\% of cases it is 2 times smaller. Only in a few percent of cases ({located in the magnetic knots relative to the observer}) the flux can be enhanced 10 times and more $F/F_\mathrm{av}>10$. On the other hand in up to 55 \% of {the} cases {the} flux can be reduced in 10 times $F/F_\mathrm{av}<10$ in {the magnetic} voids. This mean{s} that for most of {the} sources {the} UHECR flux will stay same or will be reduced, while only for {a} small fraction of sources it will be significantly increased, giving {the} possibility to observe {a} 'hot spot' in the UHECR spectrum. {The average flux through the sphere remains the same at all distances. Particles do not interact when propagating through a magnetic field.}

In Fig. \ref{fig:KnotDist} we show {an} example of $E=10$~EV {particle} fluxes relative to average one in the directions to magnetic knot, void and filament as a function of the distance from the source. {For this figure the magnetic knot, void and filament are defined at the distance of 10 Mpc, see Fig. 6 in~\citet{dolgikh2022causticlike}.} The black solid line marks an average flux $F=F_\mathrm{av}$. One can see that in example of Fig. \ref{fig:KnotDist} filament at knot have maximal flux at 12 Mpc from source and disappeared at 25-30 Mpc distance. Moreover, filament transforms into void between 30 Mpc and 50 Mpc from source.

Finally we constructed the energy spectrum of the source for an observer located in the knot and in the {void}, see Fig. \ref{fig:KnotSpectrum}. One can see that the observed flux deviates significantly from the average one at the energies around 10-30 EeV. These deviations disappear both at low and high energies. At the highest energies $\sim$~100 EeV the deflections become too small to produce strong enough flux variations. On the other hand at low energy end of the spectrum UHECR propagation becomes diffusive which is also washing out the filamentary pattern.

\subsection{Source images}
{Although the observer is quite small, its non-zero size can cause distortion of the source image. The effect is more pronounced for the sources located closer to the observer. The particles that hit the opposite edges of the observer, will have a slightly different arrival direction as seen from the center of the sphere, which is not related to the magnetic field. We take this effect into account and compensate for it in order to exclude it from the observation picture. To do this, a small correction is added to the arrival direction of the particles that hit the observer: $ \vec p_{end} \rightarrow \vec p_{end}-\frac{\vec r_{end}}{|\vec r_{end}-\vec r_{source}|} $, where $\vec r_{end}$ is final coordinates of the particle (observer will be in (0,0,0) coordinates), $\vec r_{source}$ is coordinates of the source.} 
\begin{figure*}
    \includegraphics[width=\linewidth]{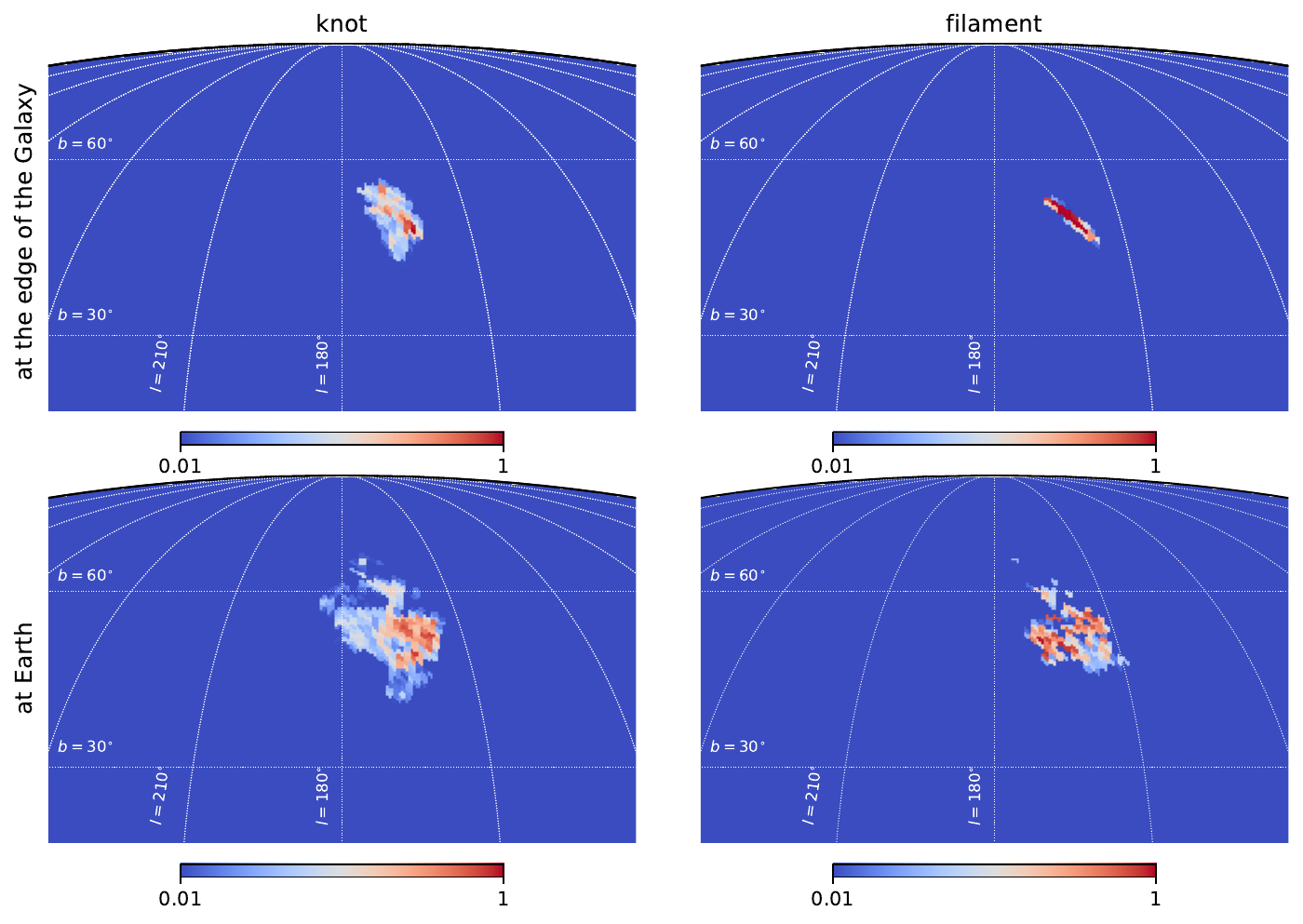}
    \caption{Source images seen by the observer located at knot and filament calculated at the edge of the Galaxy and after propagation through the GMF. Color encodes the number of hits to a given pixel. In both cases the intrinsic source luminosity is adjusted so that total number of events from the source observed at Earth matches with the cumulative number of TA events falling inside the hot spot circle of $25^\circ$ (44 events for the 15-year TA SD data \citep{Kim:2023ksw}). The blue color corresponds to the background level which is shown as uniform for clarity.}
    \label{fig:images_all}
\end{figure*}

We have studied two cases of observer location {in Fig.\ref{fig:images_all}}: observer in a filament and observer in knot. {Relative to the observer, it looks like two different environments of the extragalactic magnetic field.} We didn't consider the most likely case of the {magnetic} void (it is about $\sim 70\%$, see Fig.\ref{fig:propabilities}) as in this case the flux is extremely small and hard for detection. {The} knot case produces a bright smudged spot but this case is the most unlikely. Thus, if we have many sources of UHECR around us at a distance of tens Mpc, we see only a few of them. With respect to most sources, we will be in the voids. Sources at greater distances (more 100 Mpc) give an isotropic contribution, which decreases with distance and creates a uniform background.

On {the} top right panels of Fig.\ref{fig:images_all} {one} can see the images of the sources {at the edge of our Galaxy, i.e.} before propagation through the {Galactic} magnetic field. When {the} observer is located in a {magnetic} filament {(upper right panel)} {the} image is very stretched. {This shape of the image is not accidental and is a distinctive feature of the observer's position. The direction along which the image is stretched is always perpendicular to the magnetic filament which is the result of particle focusing. In order to have the flux enhancement in a filament, the magnetic field should bend UHECR trajectories along a direction perpendicular to the direction of the filament itself. Therefore, for an observer in the filament, particles will come from directions perpendicular to the filament. To the contrary, in the case of the observer in the magnetic knot, the image is more symmetric.} 

After passing the {Galactic} magnetic field, a smeared spots could be formed as in {the} bottom panels of Fig.\ref{fig:images_all}.  {To be more specific,} the positions of the sources in the sky were chosen in a way to correspond to {the} TA hotspot direction before GMF lensing. {The quantitative prediction will certainly depend on the specific GMF model used, but the overall picture will stay the same. The uncertainty of the UHECR deflections in the GMF was recently analysed in \citet{Unger:2023lob} (in particular, see Fig.~18 of \citet{Unger:2023lob}).}

{In order to analyze the size of the spot we have repeated the analysis performed by Telescope Array collaboration for the search of the hotspot~\citep{TelescopeArray:2014tsd} with the event set randomly generated using the density seen by the observer located at knot and filament as shown in the bottom panels of Fig.\ref{fig:images_all}. We have found that the maximum Li-Ma statistics in both cases corresponds to the 15 degrees oversampling radius, which is of the same order of magnitude as the sizes of known hotspots in the UHECR data. The hotspot observation is consistent with the discussed scenario. At the same time, under the latter scenario, some of the sources fall into the void, which requires an increased overall source density. }

We performed an analysis {for the particles of the rigidity} 10-20 EV. Since UHECRs at {the} highest energies can be dominated by intermediate nuclei according to {the} Auger data, propagation of {particles of those rigidities} would correspond to propagation of nuclei with Z times larger energies in the same magnetic fields. In particular UHECRs with $E>57$ EeV would correspond to CNO nuclei. As was shown for example in \citet{Neronov:2021xua}, CNO nuclei of required energies can reach us from up to 70 Mpc distance.

\section{Conclusions}

In this paper we studied {the} influence of the intergalactic and {Galactic} magnetic fields on the images of nearby sources in the UHECR data. In particular we studied {the} possibility {of explaining the} hot spots in UHECR data by {the} contribution of nearby sources. In our previous paper~\citet{dolgikh2022causticlike} we have shown that the effects of UHECR propagation in a turbulent intergalactic magnetic field in the small-angle scattering regime produce {a} non-trivial caustic-like pattern in the {spatial} density of cosmic rays, with {magnetic} knots, filaments and voids in the density profile at {a} given distance from {the} source. Here we studied {the} dependence of {the} density profile from {the} distance from {the} source and from energy for fixed properties of {the} magnetic field with {a} strength 1~nG and {a} coherence length {of} 1~Mpc. We found that for {a} large range of distances and energies of UHECR {, the} caustic-like pattern stays similar in the sky. This means that for {a} given position of {the} Milky Way galaxy in this pattern we get (de-)amplification of original flux of {the} source.  

Final source images after propagation in the {Galactic} magnetic field are shown in Fig. \ref{fig:images_all}. One can see that both {magnetic} knot and filament structures can provide final hot spots in the UHECR image, while location in the {magnetic} void significantly reduce {the} contribution of {a} given source to the observed UHECR flux. 

For {a} given class of UHECR sources we expect that most of {the} nearby sources are located in the {magnetic} voids and we see only few of sources, which are located in the filaments. With {a} small probability one can see even brighter amplification of {the} sources's flux in the knot. 

Thus, in this work we show how hot spots in UHECR data can form. In Fig.\ref{fig:images_all} we have shown examples of {the} source images in the direction of {the} TA hot spot. We show that after propagation in {the} Galactic and intergalactic magnetic fields {the} source image can look like {the} one observed in {the} UHECR data. We performed {an} analysis for 10~EV {particles}, which would correspond to CNO nuclei for energies of {the} hot spot in {the} TA and Auger data $E>57$ EeV. Source location at several tens of Mpc is still {possible} for CNO nuclei of those energies.

\paragraph{Acknowledgements.}  Modeling of propagation of high-energy radiation in the Universe has been supported by the Russian Science Foundation grant 22-12-00253 (K.D. and G.R.). The work of K.D. has been supported by the fellowship of Basis Foundation (grant No. 22-1-5-92-1). The work of A.K. has been supported by the IISN project No. 4.4501.18. Some of the results in this paper have been derived using the healpy and HEALPix\footnote{http://healpix.sf.net}\citep{2005ApJ...622..759G} packages.

\bibliographystyle{jasr-model5-names}
\biboptions{authoryear}
\bibliography{refs_img}

\begin{thebibliography}{28}
\expandafter\ifx\csname natexlab\endcsname\relax\def\natexlab#1{#1}\fi
\ifx\xfnm\relax \def\xfnm[#1]{\unskip,\space#1}\fi

\bibitem[{Aab et~al.(2020)}]{PierreAuger:2020kuy}
\bibinfo{author}{Aab, A.} et~al. (\bibinfo{collaboration}{Pierre Auger}) (\bibinfo{year}{2020}).
\newblock \bibinfo{title}{{Features of the Energy Spectrum of Cosmic Rays above 2.5\texttimes{}10$^{18}$ eV Using the Pierre Auger Observatory}}.
\newblock {\it \bibinfo{journal}{Phys. Rev. Lett.}\/},  {\it \bibinfo{volume}{125}\/}\bibinfo{issue}{(12)}, \bibinfo{pages}{121106}. \DOIprefix\doi{10.1103/PhysRevLett.125.121106}. \href{http://arxiv.org/abs/2008.06488}{\tt arXiv:2008.06488}.

\bibitem[{Abbasi et~al.(2008)}]{HiRes:2007lra}
\bibinfo{author}{Abbasi, R.~U.} et~al. (\bibinfo{collaboration}{HiRes}) (\bibinfo{year}{2008}).
\newblock \bibinfo{title}{{First observation of the Greisen-Zatsepin-Kuzmin suppression}}.
\newblock {\it \bibinfo{journal}{Phys. Rev. Lett.}\/},  {\it \bibinfo{volume}{100}\/}, \bibinfo{pages}{101101}. \DOIprefix\doi{10.1103/PhysRevLett.100.101101}. \href{http://arxiv.org/abs/astro-ph/0703099}{\tt arXiv:astro-ph/0703099}.

\bibitem[{Abbasi et~al.(2014)}]{TelescopeArray:2014tsd}
\bibinfo{author}{Abbasi, R.~U.} et~al. (\bibinfo{collaboration}{Telescope Array}) (\bibinfo{year}{2014}).
\newblock \bibinfo{title}{{Indications of Intermediate-Scale Anisotropy of Cosmic Rays with Energy Greater Than 57 EeV in the Northern Sky Measured with the Surface Detector of the Telescope Array Experiment}}.
\newblock {\it \bibinfo{journal}{Astrophys. J. Lett.}\/},  {\it \bibinfo{volume}{790}\/}, \bibinfo{pages}{L21}. \DOIprefix\doi{10.1088/2041-8205/790/2/L21}. \href{http://arxiv.org/abs/1404.5890}{\tt arXiv:1404.5890}.

\bibitem[{Abraham et~al.(2008)}]{PierreAuger:2008rol}
\bibinfo{author}{Abraham, J.} et~al. (\bibinfo{collaboration}{Pierre Auger}) (\bibinfo{year}{2008}).
\newblock \bibinfo{title}{{Observation of the suppression of the flux of cosmic rays above $4\times 10^{19}$eV}}.
\newblock {\it \bibinfo{journal}{Phys. Rev. Lett.}\/},  {\it \bibinfo{volume}{101}\/}, \bibinfo{pages}{061101}. \DOIprefix\doi{10.1103/PhysRevLett.101.061101}. \href{http://arxiv.org/abs/0806.4302}{\tt arXiv:0806.4302}.

\bibitem[{Abu-Zayyad et~al.(2013)}]{TelescopeArray:2012qqu}
\bibinfo{author}{Abu-Zayyad, T.} et~al. (\bibinfo{collaboration}{Telescope Array}) (\bibinfo{year}{2013}).
\newblock \bibinfo{title}{{The Cosmic Ray Energy Spectrum Observed with the Surface Detector of the Telescope Array Experiment}}.
\newblock {\it \bibinfo{journal}{Astrophys. J. Lett.}\/},  {\it \bibinfo{volume}{768}\/}, \bibinfo{pages}{L1}. \DOIprefix\doi{10.1088/2041-8205/768/1/L1}. \href{http://arxiv.org/abs/1205.5067}{\tt arXiv:1205.5067}.

\bibitem[{Alves~Batista et~al.(2016)Alves~Batista, Dundovic, Erdmann, Kampert, Kuempel, M\"uller, Sigl, van Vliet, Walz \& Winchen}]{AlvesBatista:2016vpy}
\bibinfo{author}{Alves~Batista, R.}, \bibinfo{author}{Dundovic, A.}, \bibinfo{author}{Erdmann, M.} et~al. (\bibinfo{year}{2016}).
\newblock \bibinfo{title}{{CRPropa 3 - a Public Astrophysical Simulation Framework for Propagating Extraterrestrial Ultra-High Energy Particles}}.
\newblock {\it \bibinfo{journal}{JCAP}\/},  {\it \bibinfo{volume}{05}\/}, \bibinfo{pages}{038}. \DOIprefix\doi{10.1088/1475-7516/2016/05/038}. \href{http://arxiv.org/abs/1603.07142}{\tt arXiv:1603.07142}.

\bibitem[{Alves~Batista et~al.(2022)}]{AlvesBatista:2022vem}
\bibinfo{author}{Alves~Batista, R.} et~al. (\bibinfo{year}{2022}).
\newblock \bibinfo{title}{{CRPropa 3.2 \textemdash{} an advanced framework for high-energy particle propagation in extragalactic and galactic spaces}}.
\newblock {\it \bibinfo{journal}{JCAP}\/},  {\it \bibinfo{volume}{09}\/}, \bibinfo{pages}{035}. \DOIprefix\doi{10.1088/1475-7516/2022/09/035}. \href{http://arxiv.org/abs/2208.00107}{\tt arXiv:2208.00107}.

\bibitem[{Dolgikh et~al.(2023)Dolgikh, Korochkin, Rubtsov, Semikoz \& Tkachev}]{dolgikh2022causticlike}
\bibinfo{author}{Dolgikh, K.}, \bibinfo{author}{Korochkin, A.}, \bibinfo{author}{Rubtsov, G.} et~al. (\bibinfo{year}{2023}).
\newblock \bibinfo{title}{{Caustic-Like Structures in UHECR Flux after Propagation in Turbulent Intergalactic Magnetic Fields}}.
\newblock {\it \bibinfo{journal}{J. Exp. Theor. Phys.}\/},  {\it \bibinfo{volume}{136}\/}\bibinfo{issue}{(6)}, \bibinfo{pages}{704--710}. \DOIprefix\doi{10.1134/S1063776123060031}. \href{http://arxiv.org/abs/2212.01494}{\tt arXiv:2212.01494}.

\bibitem[{Eichmann et~al.(2022)Eichmann, Kachelrie\ss{} \& Oikonomou}]{Eichmann:2022ias}
\bibinfo{author}{Eichmann, B.}, \bibinfo{author}{Kachelrie\ss{}, M.},  \& \bibinfo{author}{Oikonomou, F.} (\bibinfo{year}{2022}).
\newblock \bibinfo{title}{{Explaining the UHECR spectrum, composition and large-scale anisotropies with radio galaxies}}.
\newblock {\it \bibinfo{journal}{JCAP}\/},  {\it \bibinfo{volume}{07}\/}\bibinfo{issue}{(07)}, \bibinfo{pages}{006}. \DOIprefix\doi{10.1088/1475-7516/2022/07/006}. \href{http://arxiv.org/abs/2202.11942}{\tt arXiv:2202.11942}.

\bibitem[{Giacalone \& Jokipii(1999)}]{Giacalone_1999}
\bibinfo{author}{Giacalone, J.},  \& \bibinfo{author}{Jokipii, J.~R.} (\bibinfo{year}{1999}).
\newblock \bibinfo{title}{The transport of cosmic rays across a turbulent magnetic field}.
\newblock {\it \bibinfo{journal}{Astrophys. J.}\/},  {\it \bibinfo{volume}{520}\/}\bibinfo{issue}{(1)}, \bibinfo{pages}{204}. \URLprefix \url{https://dx.doi.org/10.1086/307452}. \DOIprefix\doi{10.1086/307452}.

\bibitem[{Giacinti et~al.(2010)Giacinti, Kachelriess, Semikoz \& Sigl}]{Giacinti:2010dk}
\bibinfo{author}{Giacinti, G.}, \bibinfo{author}{Kachelriess, M.}, \bibinfo{author}{Semikoz, D.~V.} et~al. (\bibinfo{year}{2010}).
\newblock \bibinfo{title}{{Ultrahigh Energy Nuclei in the Galactic Magnetic Field}}.
\newblock {\it \bibinfo{journal}{JCAP}\/},  {\it \bibinfo{volume}{08}\/}, \bibinfo{pages}{036}. \DOIprefix\doi{10.1088/1475-7516/2010/08/036}. \href{http://arxiv.org/abs/1006.5416}{\tt arXiv:1006.5416}.

\bibitem[{Giacinti et~al.(2011)Giacinti, Kachelriess, Semikoz \& Sigl}]{Giacinti:2011uj}
\bibinfo{author}{Giacinti, G.}, \bibinfo{author}{Kachelriess, M.}, \bibinfo{author}{Semikoz, D.~V.} et~al. (\bibinfo{year}{2011}).
\newblock \bibinfo{title}{{Ultrahigh Energy Nuclei in the Turbulent Galactic Magnetic Field}}.
\newblock {\it \bibinfo{journal}{Astropart. Phys.}\/},  {\it \bibinfo{volume}{35}\/}, \bibinfo{pages}{192--200}. \DOIprefix\doi{10.1016/j.astropartphys.2011.07.006}. \href{http://arxiv.org/abs/1104.1141}{\tt arXiv:1104.1141}.

\bibitem[{{G{\'o}rski} et~al.(2005){G{\'o}rski}, {Hivon}, {Banday}, {Wandelt}, {Hansen}, {Reinecke} \& {Bartelmann}}]{2005ApJ...622..759G}
\bibinfo{author}{{G{\'o}rski}, K.~M.}, \bibinfo{author}{{Hivon}, E.}, \bibinfo{author}{{Banday}, A.~J.} et~al. (\bibinfo{year}{2005}).
\newblock \bibinfo{title}{{HEALPix: A Framework for High-Resolution Discretization and Fast Analysis of Data Distributed on the Sphere}}.
\newblock {\it \bibinfo{journal}{\apj}\/},  {\it \bibinfo{volume}{622}\/}, \bibinfo{pages}{759--771}. \DOIprefix\doi{10.1086/427976}. \href{http://arxiv.org/abs/arXiv:astro-ph/0409513}{\tt arXiv:arXiv:astro-ph/0409513}.

\bibitem[{Greisen(1966)}]{Greisen:1966jv}
\bibinfo{author}{Greisen, K.} (\bibinfo{year}{1966}).
\newblock \bibinfo{title}{{End to the cosmic ray spectrum?}}
\newblock {\it \bibinfo{journal}{Phys. Rev. Lett.}\/},  {\it \bibinfo{volume}{16}\/}, \bibinfo{pages}{748--750}. \DOIprefix\doi{10.1103/PhysRevLett.16.748}.

\bibitem[{{Jansson} \& {Farrar}(2012)}]{JF_GMF_1}
\bibinfo{author}{{Jansson}, R.},  \& \bibinfo{author}{{Farrar}, G.~R.} (\bibinfo{year}{2012}).
\newblock \bibinfo{title}{{A New Model of the Galactic Magnetic Field}}.
\newblock {\it \bibinfo{journal}{\apj}\/},  {\it \bibinfo{volume}{757}\/}\bibinfo{issue}{(1)}, \bibinfo{pages}{14}. \DOIprefix\doi{10.1088/0004-637X/757/1/14}. \href{http://arxiv.org/abs/1204.3662}{\tt arXiv:1204.3662}.

\bibitem[{Jansson \& Farrar(2012)}]{JF_GMF_2}
\bibinfo{author}{Jansson, R.},  \& \bibinfo{author}{Farrar, G.~R.} (\bibinfo{year}{2012}).
\newblock \bibinfo{title}{{The Galactic Magnetic Field}}.
\newblock {\it \bibinfo{journal}{Astrophys. J. Lett.}\/},  {\it \bibinfo{volume}{761}\/}, \bibinfo{pages}{L11}. \DOIprefix\doi{10.1088/2041-8205/761/1/L11}. \href{http://arxiv.org/abs/1210.7820}{\tt arXiv:1210.7820}.

\bibitem[{Jasche et~al.(2020)Jasche, van Vliet \& Rachen}]{CRPropaLearns}
\bibinfo{author}{Jasche, J.}, \bibinfo{author}{van Vliet, A.},  \& \bibinfo{author}{Rachen, J.~P.} (\bibinfo{year}{2020}).
\newblock \bibinfo{title}{{Targeting Earth: CRPropa learns to aim}}.
\newblock {\it \bibinfo{journal}{PoS}\/},  {\it \bibinfo{volume}{ICRC2019}\/}, \bibinfo{pages}{447}. \DOIprefix\doi{10.22323/1.358.0447}. \href{http://arxiv.org/abs/1911.05048}{\tt arXiv:1911.05048}.

\bibitem[{Jedamzik \& Saveliev(2019)}]{Jedamzik:2018itu}
\bibinfo{author}{Jedamzik, K.},  \& \bibinfo{author}{Saveliev, A.} (\bibinfo{year}{2019}).
\newblock \bibinfo{title}{{Stringent Limit on Primordial Magnetic Fields from the Cosmic Microwave Background Radiation}}.
\newblock {\it \bibinfo{journal}{Phys. Rev. Lett.}\/},  {\it \bibinfo{volume}{123}\/}\bibinfo{issue}{(2)}, \bibinfo{pages}{021301}. \DOIprefix\doi{10.1103/PhysRevLett.123.021301}. \href{http://arxiv.org/abs/1804.06115}{\tt arXiv:1804.06115}.

\bibitem[{Kalashev et~al.(2023)Kalashev, Korochkin, Neronov \& Semikoz}]{Kalashev_2023}
\bibinfo{author}{Kalashev, O.}, \bibinfo{author}{Korochkin, A.}, \bibinfo{author}{Neronov, A.} et~al. (\bibinfo{year}{2023}).
\newblock \bibinfo{title}{{Modeling the propagation of very-high-energy \ensuremath{\gamma}-rays with the CRbeam code: Comparison with CRPropa and ELMAG codes}}.
\newblock {\it \bibinfo{journal}{Astron. Astrophys.}\/},  {\it \bibinfo{volume}{675}\/}, \bibinfo{pages}{A132}. \DOIprefix\doi{10.1051/0004-6361/202243364}. \href{http://arxiv.org/abs/2201.03996}{\tt arXiv:2201.03996}.

\bibitem[{Kampert et~al.(2012)Kampert, Watson \& Watson}]{Kampert:2012vi}
\bibinfo{author}{Kampert, K.-H.}, \bibinfo{author}{Watson, A.~A.},  \& \bibinfo{author}{Watson, A.~A.} (\bibinfo{year}{2012}).
\newblock \bibinfo{title}{{Extensive Air Showers and Ultra High-Energy Cosmic Rays: A Historical Review}}.
\newblock {\it \bibinfo{journal}{Eur. Phys. J. H}\/},  {\it \bibinfo{volume}{37}\/}, \bibinfo{pages}{359--412}. \DOIprefix\doi{10.1140/epjh/e2012-30013-x}. \href{http://arxiv.org/abs/1207.4827}{\tt arXiv:1207.4827}.

\bibitem[{Katz et~al.(2021)}]{Katz:2021iou}
\bibinfo{author}{Katz, H.} et~al. (\bibinfo{year}{2021}).
\newblock \bibinfo{title}{{Introducing SPHINX-MHD: the impact of primordial magnetic fields on the first galaxies, reionization, and the global 21-cm signal}}.
\newblock {\it \bibinfo{journal}{Mon. Not. Roy. Astron. Soc.}\/},  {\it \bibinfo{volume}{507}\/}\bibinfo{issue}{(1)}, \bibinfo{pages}{1254--1282}. \DOIprefix\doi{10.1093/mnras/stab2148}. \href{http://arxiv.org/abs/2101.11624}{\tt arXiv:2101.11624}.

\bibitem[{Kim et~al.(2023)Kim, Ivanov, Kawata, Sagawa \& Thomson}]{Kim:2023ksw}
\bibinfo{author}{Kim, J.}, \bibinfo{author}{Ivanov, D.}, \bibinfo{author}{Kawata, K.} et~al. (\bibinfo{collaboration}{Telescope Array}) (\bibinfo{year}{2023}).
\newblock \bibinfo{title}{{Anisotropies in the arrival direction distribution of ultra-high energy cosmic rays measured by the Telescope Array surface detector}}.
\newblock {\it \bibinfo{journal}{PoS}\/},  {\it \bibinfo{volume}{ICRC2023}\/}, \bibinfo{pages}{244}. \DOIprefix\doi{10.22323/1.444.0244}.

\bibitem[{Mtchedlidze et~al.(2022)Mtchedlidze, Dom\'\i{}nguez-Fern\'andez, Du, Brandenburg, Kahniashvili, O'Sullivan, Schmidt \& Br\"uggen}]{Mtchedlidze:2021bfy}
\bibinfo{author}{Mtchedlidze, S.}, \bibinfo{author}{Dom\'\i{}nguez-Fern\'andez, P.}, \bibinfo{author}{Du, X.} et~al. (\bibinfo{year}{2022}).
\newblock \bibinfo{title}{{Evolution of Primordial Magnetic Fields during Large-scale Structure Formation}}.
\newblock {\it \bibinfo{journal}{Astrophys. J.}\/},  {\it \bibinfo{volume}{929}\/}\bibinfo{issue}{(2)}, \bibinfo{pages}{127}. \DOIprefix\doi{10.3847/1538-4357/ac5960}. \href{http://arxiv.org/abs/2109.13520}{\tt arXiv:2109.13520}.

\bibitem[{Neronov et~al.(2023)Neronov, Semikoz \& Kalashev}]{Neronov:2021xua}
\bibinfo{author}{Neronov, A.}, \bibinfo{author}{Semikoz, D.},  \& \bibinfo{author}{Kalashev, O.} (\bibinfo{year}{2023}).
\newblock \bibinfo{title}{{Limit on the intergalactic magnetic field from the ultrahigh-energy cosmic ray hotspot in the Perseus-Pisces region}}.
\newblock {\it \bibinfo{journal}{Phys. Rev. D}\/},  {\it \bibinfo{volume}{108}\/}\bibinfo{issue}{(10)}, \bibinfo{pages}{103008}. \DOIprefix\doi{10.1103/PhysRevD.108.103008}. \href{http://arxiv.org/abs/2112.08202}{\tt arXiv:2112.08202}.

\bibitem[{Pshirkov et~al.(2016)Pshirkov, Tinyakov \& Urban}]{Pshirkov:2015tua}
\bibinfo{author}{Pshirkov, M.~S.}, \bibinfo{author}{Tinyakov, P.~G.},  \& \bibinfo{author}{Urban, F.~R.} (\bibinfo{year}{2016}).
\newblock \bibinfo{title}{{New limits on extragalactic magnetic fields from rotation measures}}.
\newblock {\it \bibinfo{journal}{Phys. Rev. Lett.}\/},  {\it \bibinfo{volume}{116}\/}\bibinfo{issue}{(19)}, \bibinfo{pages}{191302}. \DOIprefix\doi{10.1103/PhysRevLett.116.191302}. \href{http://arxiv.org/abs/1504.06546}{\tt arXiv:1504.06546}.

\bibitem[{Unger \& Farrar(2023)}]{Unger:2023lob}
\bibinfo{author}{Unger, M.},  \& \bibinfo{author}{Farrar, G.~R.} (\bibinfo{year}{2023}).
\newblock \bibinfo{title}{{The Coherent Magnetic Field of the Milky Way}}.
\newblock {\it \bibinfo{journal}{Accepted for publication in \apj}\/}, . \href{http://arxiv.org/abs/2311.12120}{\tt arXiv:2311.12120}.

\bibitem[{Yüksel et~al.(2012)Yüksel, Stanev, Kistler \& Kronberg}]{Y_ksel_2012}
\bibinfo{author}{Yüksel, H.}, \bibinfo{author}{Stanev, T.}, \bibinfo{author}{Kistler, M.~D.} et~al. (\bibinfo{year}{2012}).
\newblock \bibinfo{title}{{THE} {CENTAURUS} a {ULTRAHIGH}-{ENERGY} {COSMIC}-{RAY} {EXCESS} {AND} {THE} {LOCAL} {EXTRAGALACTIC} {MAGNETIC} {FIELD}}.
\newblock {\it \bibinfo{journal}{Astrophys. J.}\/},  {\it \bibinfo{volume}{758}\/}\bibinfo{issue}{(1)}, \bibinfo{pages}{16}. \URLprefix \url{https://doi.org/10.1088/0004-637x/758/1/16}. \DOIprefix\doi{10.1088/0004-637x/758/1/16}.

\bibitem[{Zatsepin \& Kuzmin(1966)}]{Zatsepin:1966jv}
\bibinfo{author}{Zatsepin, G.~T.},  \& \bibinfo{author}{Kuzmin, V.~A.} (\bibinfo{year}{1966}).
\newblock \bibinfo{title}{{Upper limit of the spectrum of cosmic rays}}.
\newblock {\it \bibinfo{journal}{JETP Lett.}\/},  {\it \bibinfo{volume}{4}\/}, \bibinfo{pages}{78--80}.
\newblock \bibinfo{note}{[Pisma Zh. Eksp. Teor. Fiz.4,114(1966)]}.

\end{thebibliography}
\end{document}